\documentstyle[12pt]{article}
\begin{document}
\baselineskip=14pt

\newcommand{\D}{\displaystyle} 
\newcommand{\T}{\textstyle} 
\newcommand{\SC}{\scriptstyle} 
\newcommand{\SSC}{\scriptscriptstyle} 

\hyphenation{mono-chro-matic  sour-ces  Wein-berg
chang-es Strah-lung dis-tri-bu-tion com-po-si-tion elec-tro-mag-ne-tic
ex-tra-galactic ap-prox-i-ma-tion nu-cle-o-syn-the-sis re-spec-tive-ly
su-per-nova su-per-novae su-per-nova-shocks con-vec-tive down-wards
es-ti-ma-ted frag-ments grav-i-ta-tion-al-ly el-e-ments me-di-um
ob-ser-va-tions tur-bul-ence sec-ond-ary in-ter-action
in-ter-stellar spall-ation ar-gu-ment de-pen-dence sig-nif-i-cant-ly
in-flu-enc-ed par-ti-cle sim-plic-i-ty nu-cle-ar smash-es iso-topes
in-ject-ed in-di-vid-u-al nor-mal-iza-tion lon-ger con-stant
sta-tion-ary sta-tion-ar-i-ty spec-trum pro-por-tion-al cos-mic
re-turn ob-ser-va-tion-al es-ti-mate switch-over grav-i-ta-tion-al
super-galactic com-po-nent com-po-nents prob-a-bly cos-mo-log-ical-ly
Kron-berg}
\def\simle{\lower 2pt \hbox {$\buildrel < \over {\scriptstyle \sim }$}}
\def\simge{\lower 2pt \hbox {$\buildrel > \over {\scriptstyle \sim }$}}


\begin{center}
{\large {{\bf The supergalactic structure and\\
the origin of the highest energy cosmic rays}}}\\
\vskip1.0cm
Peter L. Biermann$^1$, Hyesung Kang$^{2}$, and Dongsu Ryu$^{2,3}$\\[10mm]
$^1$ Max-Planck Institute for Radioastronomy, D-53010 Bonn, Germany\\
$^2$ Department of Astronomy, University of Washigton, Seattle, WA98195, USA\\
$^3$ Dept. of Astronomy \& Space Science, Chungnam National Univ.,\\
Daejeon, Korea\\[5mm]

\end{center}


\begin{abstract}
The recent discoveries of several reliable events of high energy cosmic
rays at an energy above $10^{20}$ eV raise questions about their path
through the nearby universe.  The two analyses of, on the one hand, the
Haverah Park data set including a limited set of further events and, on
the other hand, the Akeno data set appear to have an inconsistent pattern
of arrival directions.  Both data sets showed some measure of a correlation
with the supergalactic plane, the locus of cosmologically nearby galaxies,
radio galaxies and clusters of galaxies.  In order to be able to
interprete such findings,  we need a reasonable model of the true
intergalactic magnetic field and then can expect to make further progress 
on the propagation of energetic charged particles. Using recent cosmological
simulations of structure formation in the universe, we estimate the magnetic
fields which correspond to the upper limits in the Rotation Measure to distant
radio sources.  Using the one single direct measurement of such a magnetic
field, near the Coma cluster, we thus estimate that the magnetic field strength
in supergalactic sheets and filaments may be in the range of 0.1 to 1
microgauss.  If such strengths are realized inside our Local Supercluster, this
opens up the possibility to focus charged particles in the direction
perpendicular to the supergalatic plane, analoguously but in the opposite
direction to solar wind modulation. If focusing exists, it means that for all
particles captured into the sheets, the dilution with distance $d$ is $1/d$
instead of $1/d^2$, increasing the cosmic ray flux from any source appreciably
with respect to three-dimensional expansion.  This means in effect, that we may
see sources to much larger distances than expected sofar.  This effect is
relevant only for energies for which the possible distances are smaller
than the void scale of the cosmological galaxy distribution, in the
range possibly up to 100 Mpc, but presumably less than this distance.

\end{abstract}

\section{Introduction}

The discovery of cosmic rays early this century \cite{CRA,CRB} spawned many 
observations  of these high energy particles, right up to the recent 
detection \cite{Akeno94,FE,FE95} of a significant number of particles 
beyond the energy of $10^{20}$ eV; for general reviews of these questions see,
e.g., \cite{Hillas84,Venyabook,Gaisser90,G93,TucsonCR,JPhysGCR}.  

While the search into the origin of cosmic rays still awaits the final
resolution, there are many successful steps that have been taken, from the
original Fermi-acceleration process \cite{Fermi49,Fermi54} via the argument 
that the  high energy particles ought to be extragalactic 
\cite{Cocconi56,BS87,Cocconi96}
to the more recent discoveries already mentioned.

In this brief discussion we propose to concentrate on the arrival directions 
of the most energetic cosmic rays, and their possible correlation with the 
matter distribution in the nearby universe
\cite{Fermilab,FermilabTS,PRL95,Hayashida96}.

\section{Expectations for high energy cosmic rays}

For the source region or in our Galaxy the Larmor radius $r_g$ is given by

\begin{equation}
r_g \; \approx 10  \; E_{20} \, B^{-1}_{-5} \; {\rm kpc},
\end{equation}

\noindent where $E_{20}$ is the particle energy in $10^{20}$ eV, and $B_{-5}$
is the magnetic field in units of $10^{-5}$ Gauss.

This means that the Larmor radius is larger than the thickness of the Galactic
cosmic ray disk (about a kpc; for a review of the Galactic magnetic field see
\cite{Beck96}), and of the order of the size of the source region if radio 
galaxy hot spots of very high luminosity are considered \cite{BS87}. Thus, 
at such  energies, propagation through the Galaxy is nearly in a straight line
path.

Other important limitations are obviously losses against photon or magnetic
field backgrounds, and the time required for acceleration (for a review, see
\cite{JPhysGCR}).
Detailed calculations for the propagation have been done by a variety of
scientists, e.g., Stanev \cite{Stanev96b}, and several others.

For intergalactic space the Larmor radius is conveniently scaled to
other units and can be written as

\begin{equation}
r_g \; \approx 100 \; E_{20} \, B^{-1}_{-9} \, {\rm Mpc},
\end{equation}

\noindent where the magnetic field strength is obviously given in units
of $10^{-9}$ Gauss.
This means that for the typical upper limits derived from Rotation Measure
observations (for a review see Kronberg \cite{Kronberg94}), the intergalactic
propagation is also in a nearly straight line path.
Therefore, it is meaningful to ask for the arrival directions on the sky,
and whether they correlate with any known objects or structures.

However, losses in the bath of the microwave background radiation (MBR)
limit the distance from which particles can realistically come to less
than $\simle \, 50$ to 100 Mpc.  In other words, integrating over a
presumed cosmologically homogeneous source distribution leads to
a cutoff in the summed contributions from all sources near
$5\times10^{19}$ eV.  This is the GZK cutoff, named after Greisen,
Zatsepin and Kuzmin \cite{Greisen66,ZK66,Stecker68}.

Probable sources at distances $<$ 100 Mpc are distributed in the
supergalactic plane sheet \cite{deV56,Tully86,ShaverP89,Shaver91}, which is
defined by the Local Supercluster of nearby galaxies ($<30h^{-1}$Mpc).
Therefore arrival directions should cluster toward the supergalactic
plane from energies, where MBR losses become important.
Or, in other words, from near $5\times10^{19}$ eV the arrival directions
should cluster just as the sources in our neighborhood do.
However, there should be no clustering below this energy.

\section{Test of prediction}

This prediction was made and explored in various lectures late 1994 and early
1995 \cite{ChinaCR,StockholmCR,Fermilab,FermilabTS}, and
has been tested using several data sets.

First, the Haverah Park Array data set was used, and also combined with a
fraction of other data available at the time \cite{PRL95}.  In this test the
question was asked in the following way:  Given very limited statistics, we can
test whether the arrival directions cluster better towards the Galactic plane,
or the supergalactic plane.  The measure of success was the distance of the
arrival directions to the reference plane, and the distance for homogeneously
scrambled data (in order to allow for all selection effects).  This test showed
an effect of a correlation at a level somewhat below 3 sigma.

Thus, the effect was consistent with the prediction and was visible from
$4\times 10^{19}$ eV.

Second, the Akeno Array data set was used \cite{Hayashida96}.  These authors
found another measure to be better as a test:  The distance to the Galactic and
supergalactic plane did not yield any better result than the analysis of the
Haverah Park data, but using pairs of events there appeared a tantalizing
excess of pairs lying directly on the supergalactic plane sheet.
Thus, here again the effect was consistent with the
prediction, however, using pairs of events, from $5\times10^{19}$ eV.

Third, as presented at this meeting, the combination of all events and again
using pairs and triplets of events suggested that the supergalactic plane is
the region of origin of ultrahigh energy cosmic ray events, for a fraction of
maybe 15 - 20 \%, or possibly more, of all events beyond $5\times10^{19}$ eV.

However, as pointed out by Waxman et al. \cite{Waxman96}, not only is there a
seeming inconsistency between the first two analyses,
but there is also a deeper
difficulty:  When one uses the actual distribution of galaxies farther
beyond the Local Supercluster as a measure of possible source
directions and distances, the supergalactic plane
is not such a good approximation anymore,
and so one would not really simply expect a direct
straightforward correlation.
The observed correlation is apparently better with the geometric sheet
corresponding naively to the cosmologically nearby galaxy population, rather 
than with the actual large-scale galaxy distribution.

\section{The supergalactic Plane}

Galaxies are distributed in the observed universe in a non-homogeneous pattern,
in what may loosely be described as a network of filamentary superclusters
encompassing voids, with typical void scales of (30-100) $h^{-1}$ Mpc
\cite{Shectman96,Einasto96}.
We live in one of the superclusters, namely the Local Supercluster
\cite{Tully86}.  It is a flattened condensation of nearby galaxies
centered at the Virgo cluster extending to 30$h^{-1}$ Mpc with a scale
height of 5$h^{-1}$ Mpc.
It is also connected to several nearby superclusters by filaments of
galaxies and clusters.
This observed structure of supercluster-void networks can now be well
simulated in large computers and also be interpreted \cite{Bond96}:
The large scale structure forms as the result
of gravitational instability, and then the matter flows into the
potential wells, into the sheets, filaments, where sheets intersect, and
nodes, where filaments intersect.  This means that there is baryonic
accretion flow towards the nonlinear structures in cosmological
structure formation.
The velocity of this accretion flow can be as large as about 1000 km/sec,
independent of the Hubble constant \cite{RyuKang96}.

As a consequence of the accretion flow, the cosmological magnetic field
is expected to lie mostly along sheets and filaments.
Assuming uniformity, Kronberg et al. had derived an upper limit for the true
intergalactic magnetic field of $B \; < \; 10^{-9} \; L_{rev,Mpc}^{-1/2}$
Gauss; here $L_{rev,Mpc}$ is the reversal scale of the magnetic field.
Allowing for the correlation of magnetic field to cosmic structures
we can rederive this limit and it
transforms an upper limit to $B \; < \; 10^{-6 \pm 0.5}$ Gauss along sheets
and filaments \cite{Kulsrud96,RyuKangBiermann96}.
Interestingly there is confirmation of a definitive strength of such a
magnetic field in one case, in the plane of the Coma/A1367 supercluster
\cite{Kim89}. This supercluster is about 90$h^{-1}$Mpc away from the edge 
of the Local Supercluster and they are connected by filaments of galaxies.

\vspace{.5cm}
\noindent
%
{Figure 1. A two-dimensional cut in a simulation of the evolution of
the cosmological flow in a standard cold dark matter (SCDM) universe
with total $\Omega=1$ and baryonic $\Omega_b=0.06$.
The calculation has been done in a box of $(32 h^{-1}{\rm Mpc})^3$ volume,
and the plot includes a region of $(16h^{-1}{\rm Mpc})^2$ with
a thickness of $0.25 h^{-1}{\rm Mpc}$.
The first panel shows density contours, the second panel shows
velocity vectors, and the third panel shows magnetic field vectors.
In the third panel, the vector length is proportional to the log of
magnetic field strength.}
\vspace{.5cm}

\section{Supergalactic Modulation and Confinement}

Suppose we approximate the Local Supercluster as a cosmological sheet
bounded by two plane parallel accretion shocks.
Then in the accretion flow and in the sheet we can
again write the Larmor radius

\begin{equation}
r_{g \perp}\; \approx 0.05 \; E_{19.7} \, B^{-1}_{-6} \,
(\frac{p_{\perp}}{p_{tot}})\; {\rm Mpc},
\end{equation}

\noindent and notice that the Larmor radius for the highest energy particles
is smaller than the thickness of the supergalactic sheets for 
$B_{-6} \, \simge \, 0.02 \, h \, E_{19.7} \, {p_{\perp}}/{p_{tot}}$.  This
means a rather small strength of the magnetic field may be sufficient to
contain high energy particles in the sheets.

Then the question arises : Is modulation possible in
the accretion flow analogous
to solar wind modulation of cosmic rays?
Of course, if such a modulation were possible, it would only
pertain to the momentum component perpendicular to the sheet.

The transport equation for energetic particles with $z$ along
$\perp$-direction
to the sheet can be written as

\begin{equation}
{F_{\perp}} \; = \; N \, {v_{\perp}} \, - \, \kappa_{\perp} \,
\frac{\partial}{\partial z} \, N .
\end{equation}
\begin{equation}
N \; = \; N_o \, \exp[ - \frac{\mid v_{\perp} \mid}{\kappa_{\perp}} \, z]
\end{equation}

\noindent Note ${v_{\perp}} \, < 0$ for $z$ positive,
where $v_{\perp}$ is the accretion velocity perpendicular to the sheet, and 
$\kappa_{\perp}$ the transport coefficient perpendicular to the sheet.

The critical question is then:  What numerical value can $\kappa_{\perp}$
possibly have?
The transport coefficient can be written as (the characteristic length scale)
$\times$ (the characteristic velocity scale).
If all the magnetic field is perpendicular to the sheet,
then the Larmor radius and the speed of particle, $c$, are the relevant scales,
then we have a very large $\kappa_{\perp}$,  and no modulation is
effectively possible.
But if the magnetic field has large parallel components, then convective
turbulence is dominant probably.
It means that velocity and length scales of transport are of the order of the
accretion velocity ($v_{\perp}$) to the supergalactic plane sheet and the
scale height of sheet ($H_{sgp})$, respectively.
Then for particles with $r_{g \perp} \,<H_{sgp}/2 \sim 2.5
\,h^{-1}{\rm Mpc}$ modulation is possible.
Thus, if supergalactic modulation can exist, then we can have
weak confinement along the supergalactic sheet.

In three-dimensional expansion $1/d^2$ dilution and a gradual weakening
of the particle flux by interaction with the MBR lead at large
distances to a total cutoff above $5\times 10^{19}$ eV.
If some fraction of the particles originated from the sources in the
Local Supercluster are confined along the sheet, we
have only a two-dimensional $1/d$ dilution but also interaction with
the MBR.  This effect is clearly relevant only beyond $5 \times
10^{19}$ eV, since at lower energies we can see sources beyond the
Local Superclusters.
In other words, for confined particles focusing along the
two-dimensional sheet is possibly
stronger than original source distribution beyond $5
\times 10^{19}$ eV.

We offer this possibility as a potential solution to the conundrum posed
by the
two data analyses about arrival directions in the literature. This is a
possible
explanation for the Haverah Park and Agasa results.

\section{Tests}

There are a number of tests that can be done in the next few years, or could be
done now:

\begin{itemize}

\item{} In such a picture we have stronger magnetic fields along
sheets/filaments of
superclusters, of order

\begin{equation}
B \; \simle \; 10^{-6 \pm 0.5} \, {\rm Gauss}
\end{equation}

Radio polarization observations of cosmologically distant radio sources
(group of Phil P. Kronberg) will provide the most stringent check.

\item{} All present and future events beyond $4 \times 10^{19}$ eV
from the various
arrays Akeno, Haverah Park, Fly's Eye, Yakutsk, Volcano Ranch, and in the
future Auger should be combined to repeat the analysis:
At this meeting there was a first
report of such an attempt, with very interesting results.

\item{}  We need to verify specific source candidates, such as the radio
galaxies 3C134, NGC315, and M87.
We also suggested the accretion shocks around the large scale structure
as a possible candidate \cite{Kang95,Kang96}.

\end{itemize}

The future of our attempts to understand the origin of these very high energy
particles promises to be challenging.

\section{Acknowledgement}

The report was inspired by PLB's interactions
and collaborations with Rainer Beck, Venya Berezinsky,
Jim Cronin, Torsten En{\ss}lin, Tom Gaisser,
Gopal-Krishna, Tom Kephardt, Phil Kronberg,
Hinrich Meyer,
Motohiko Nagano, Michal Ostrowski, Ray Protheroe, J{\"o}rg Rachen,
Todor Stanev, Alan Watson, Tom Weiler, Xiang-Ping Wu, and Gaurang Yodh.
The work by DR was supported in part by Seoam Scholarship Foundation.

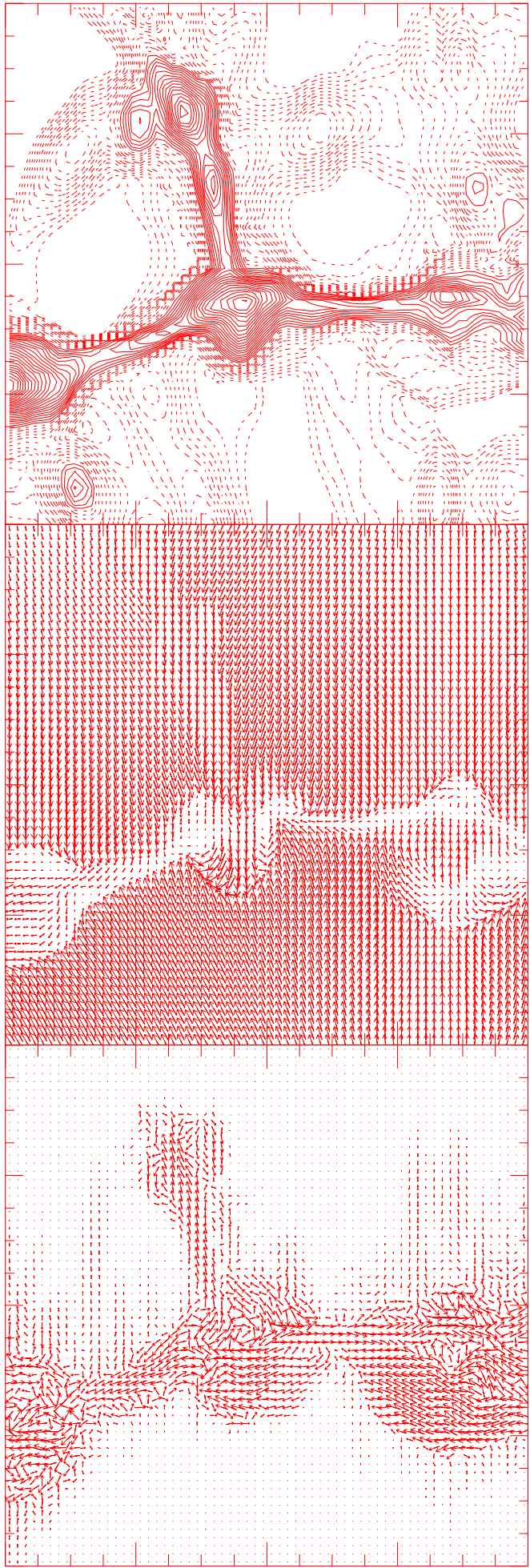
\end{document}